\documentclass[fleqn,onecolumn,11pt]{wlscirep}

\usepackage[utf8]{inputenc}
\usepackage[T1]{fontenc}
\usepackage{dsfont}

\usepackage[normalem]{ulem}

\title{Three-body bound states in antiferromagnetic spin ladders}

\author[1]{Gary Schmiedinghoff}
\author[1]{Leanna M\"uller}
\author[2]{Umesh Kumar}
\author[1]{G\"otz S. Uhrig}
\author[2,3,*]{Benedikt Fauseweh}

\affil[1]{Condensed Matter Theory, TU Dortmund University,
Otto-Hahn-Straße 4, 44227 Dortmund, Germany}
\affil[2]{Theoretical Division, 
Los Alamos National Laboratory, Los Alamos, New Mexico 87545, USA}
\affil[3]{Institute for Software Technology, German Aerospace Center (DLR), 
Linder H\"ohe, 51147 Cologne, Germany}

\affil[*]{benedikt.fauseweh@dlr.de}

\begin{abstract}
Stable bound quantum states are ubiquitous in nature. 
Mostly, they result from the interaction of only pairs of particles, so called two-body interactions,
even when large complex many-particle structures are formed. We show
that three-particle bound states occur in a generic, experimentally accessible solid
state system: antiferromagnetic spin ladders, related to 
high-temperature superconductors.
This binding is induced by genuine three-particle interactions;
without them there is no bound state. We compute the 
dynamic exchange structure factor required for the experimental
detection of the predicted state by resonant inelastic
X-ray scattering for realistic material parameters. Our work enables us to quantify these elusive interactions and unambiguously establishes their effect on the dynamics of the quantum many-particle state.
\end{abstract}
\begin{document}

\flushbottom
\maketitle

\thispagestyle{empty}

\section*{Introduction}
Binding phenomena were at the origin of the development of quantum mechanics. 
The quest to understand why electrons flying around positive nuclei form 
stable entities, namely atoms, has led to a completely new foundation of physics 
over hundred years ago. The formation of the structures from nuclei and electrons over
 atoms, small and large molecules to macroscopic liquids and solids is founded on 
quantum mechanical binding. But also down to smaller and smaller length scales binding is crucial in the formation of 
the atomic nuclei formed by protons and neutrons due to the strong force and
of heavy particles in the Standard Model of high-energy physics
leading to hadrons, namely  mesons, bound states of two elementary quarks, and baryons, bound states of three elementary quarks.

Binding effects are especially dominant in correlated antiferromagnetic spin system in low
dimensions. Such systems have already revealed some fascinating binding phenomena before. 
A chain of localized spins $S=1/2$ coupled on adjacent sites
and subjected to a magnetic field displays fractional excitations ((anti-)psinons) of half-integer quantum numbers, spin flips (magnons)
with integer spin number, and in particular bound states of $n$ magnons ($n$-strings).
These states were theoretically predicted up to 90 years ago
\cite{bethe31,gaudi71,takah72,kohno09}, but have been verified experimentally only recently \cite{Loidl2018N,PhysRevLett.123.067202}. Spin ladders, in particular, are solid state systems which are intensively studied due to their relation to high-temperature superconductors\cite{dagot99}.
Moreover, they represent a paradigm for an entangled magnetic many-body systems without long-range order \cite{dagot96}. Two-body bound states in spin ladders have been measured for $S=0$ mediated by a phonon in infrared absorption \cite{PhysRevLett.87.127002} and for $S=1$ by inelastic neutron scattering \cite{PhysRevLett.98.027403}.

In spite of the abundance and complexity of bound states most of the binding  phenomena are induced by two-body interactions. This means that the underlying potential only depends on the positions and properties of pairs of particles. 
This is particularly obvious in chemistry where all atoms and molecules are held together essentially by Coulomb potentials. 
It is the paradigm of a two-body interaction which can bind not only two, but almost infinitely many particles. 

In this communication, we show that there are also existing, realistic
solid state systems where a genuine (irreducible) three-body
interaction provides the vital extra for the formation of a bound state. This bound state would not exist if only two-body interactions were present. The key point is that
an irreducible three-body interaction acts only if three particles are present. It has no effect on two particles or a single particle. We explicitly calculate the effect of this bound state for dynamical correlations in spin ladders using continuous unitary transformations (CUT). We find an effect stronger than $50\%$ for experimentally relevant observables that can be measured with resonant inelastic X-ray scattering (RIXS).

\section*{Results}

\subsection*{Theoretical model}

\begin{figure}
	\centering
	\includegraphics[width=1.00\linewidth]{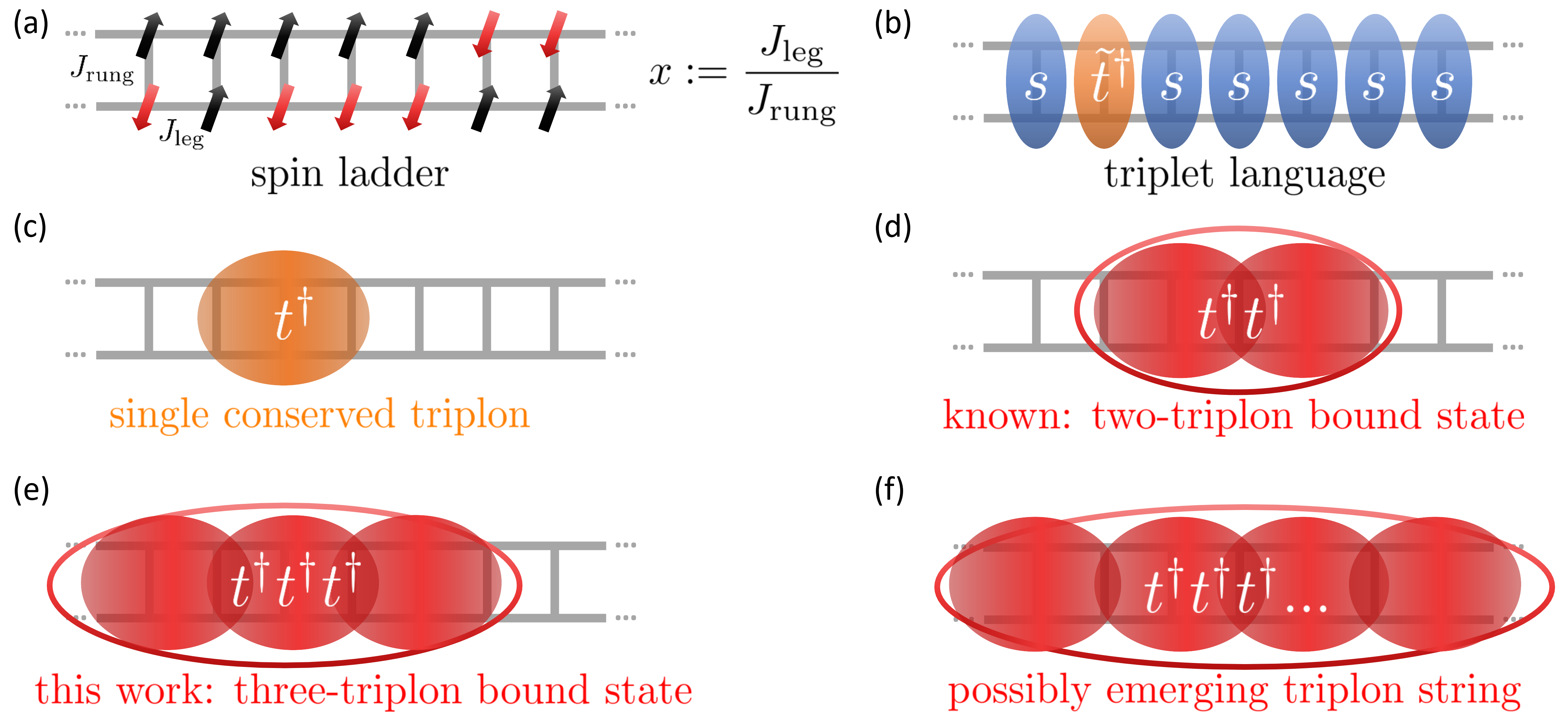}
	\caption{{Multi-triplon states in spin ladders}. (a) The spins $S=1/2$ at each
	vertex of the ladder are coupled by leg ($J_\text{leg}$),
	and rung ($J_\text{rung}$), couplings. The black arrows indicate spin-up and the red arrows indicate spin-down.
	(b)  For $J_\text{leg} = 0$, the spins on each rung 
	form singlets (blue ellipses) in the ground state and local $S=1$ triplet excitations (orange ellipse).
	(c) Non-local triplons (wide orange ellipse) are the elementary excitations in spin ladders. They exist in the $\Delta S = 1$ sector and can be detected via inelastic neutron scattering.
	(d) Two-triplon interactions lead to the formation of two-triplon bound states (red double ellipse) in the $\Delta S = 0,1$ sectors. 
	(e) Three-triplon interactions are strong enough to form three-triplon bound states in the $\Delta S = 0$ sector. 
	(f) $n$-strings of triplons can emerge; they are predicted in strongly frustrated 
	spin ladders with additional diagonal couplings (not shown) in each plaquette.}
	\label{fig:triplon_strings}
\end{figure}

We consider the  antiferromagnetic spin ladder \cite{dagot96,dagot99} 
\begin{equation}
\label{Jmodel} 
H = J_\text{rung}\sum_{i=1}^{N} \textbf{S}_{i,1}\cdot \textbf{S}_{i,
2} +J_\text{leg}\sum_{i,\tau=0,1}
\textbf{S}_{i,\tau}\cdot \textbf{S}_{i+1,\tau} 
\end{equation} 
where  $\textbf{S}_{i,\tau}$ denotes the  vector spin operator at rung $i$ along the leg and on site $\tau = 0,1$, see Fig.~\ref{fig:triplon_strings}(a). Spin ladders with $x:=J_\text{leg}/J_\text{rung}\gtrapprox 1$ are realized in cuprates to a high degree of accuracy \cite{dagot99}
so that experimental verification of our predictions is possible.

For $x = 0$, the excitations are local $S=1$ triplets above the $S=0$ singlet ground state (Fig.~\ref{fig:triplon_strings}(b)). For $x > 0$, the elementary excitation is no longer localized on one rung only, but it is smeared out over a number of rungs, the size of which is given by the correlation length (Fig.~\ref{fig:triplon_strings}(c)). It is now called a triplon \cite{trebs00,PhysRevLett.87.167204,schmi03c,schmi05b,PhysRevB.96.115150}. Note that triplon excitations are typical also in other dimerized systems, such as in Shastry-Sutherland magnets \cite{McClarty2017}. In contrast to the number of triplets, the number of triplons is conserved also for finite $x$ and therefore, triplons provide a natural basis for the description of dimerized systems \cite{Kohno2007,Sachdev2008}. For clarity, we distinguish
in this article between the initial non-conserved triplets and the conserved triplons.
In literature, this distinction is not always used \cite{Kohno2007,Sachdev2008,sachd12}.

We stress that the original triplets are not the 
appropriate elementary excitations because already in the
ground state there is an infinite number of them admixed.
Hence, it does not make sense to refer to a one-, two-
or three-triplet state since any eigen state comprises
an infinite number of them. Instead, we have to use the
elementary excitation resulting from renormalization. This is analogous to the vacuum fluctuations
in quantum field theories. The observed and measurable
elementary particles are quasiparticles, dressed by vacuum
fluctuations, whose properties stem from renormalization.

The leg coupling also leads to an attractive
interaction between pairs of triplons which are not far apart from each other,  see Fig.~\ref{fig:triplon_strings}(d).
The competition between attractive interaction and kinetic energy determines the formation of bound states in the spin ladders. As triplons are equivalent to mobile $S = 1$ spins
that interact antiferromagnetically on the spin ladder, it is expected that the energetically lowest bound state is in the total $S = 0$ sector.
Two-body bound states with total spin $S=0$ and $S=1$ 
induced by the two-triplon attraction are established theoretically \cite{trebs00,PhysRevLett.87.167204,schmi05b} and have been measured in experiments \cite{PhysRevLett.87.127002, PhysRevLett.98.027403}. The aim of the present communication is
to show that three-triplon bound states occur, see Fig.~\ref{fig:triplon_strings}(e),
which are essentially due to genuine three-triplon interactions. 

\begin{figure}
	\centering
	\includegraphics[width=0.80\linewidth]{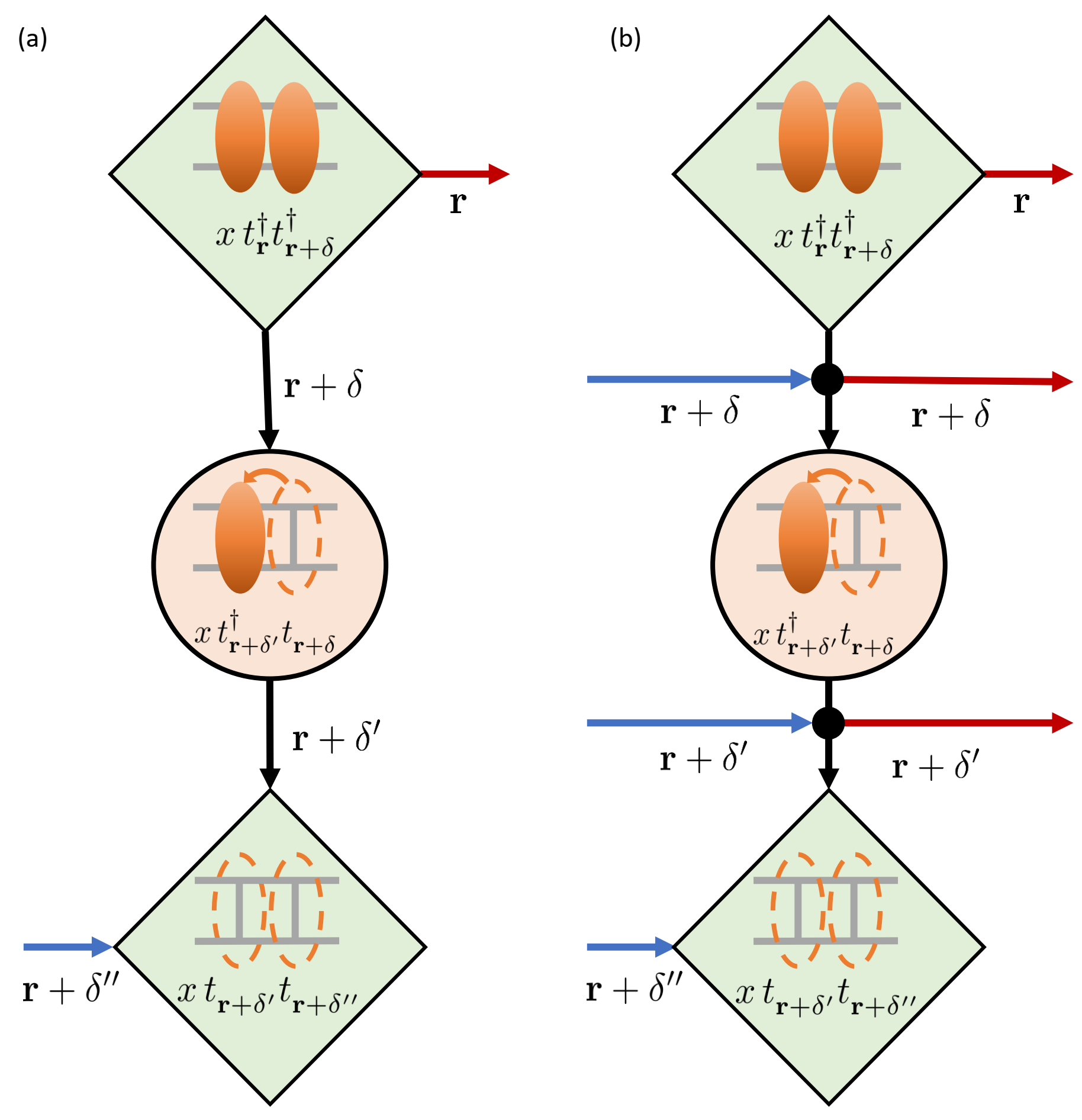}
	\caption{Origin of three-triplon interactions. The term is depicted in real space at dimer $r$ and 
	interdimer distances $\delta, \delta', \delta''$; 
	note that this term arises in any dimension
	and for any lattice model with finite dimensional local degrees of freedom.  Finite $x$ implies hopping, pair creation and annihilation processes during the renormalization by CUT. The blue arrows indicate the incoming triplons, red the scattered triplons and the black arrows internal triplon propagation. For normal bosons (a), the combined process is single-particle irreducible and corresponds to an effective hopping. For triplons (b), the hardcore constraint (black circles) induces three-triplon interactions in leading order $x^3$.
	The explicit expression for $\delta=1, \delta'=2, \delta''=3$ is given in the Supplementary Methods.} 
	\label{fig:triplon_interaction_process}
\end{figure}

The occurrence of the three-triplon interactions results from
a mechanism very analogous to the attractive interaction between 
electrons resulting from the exchange of phonons which 
eventually leads to the formation of Cooper pairs and superconductivity.
The propagation of an electron excites a phonon which is
captured by another propagating electron inducing their
attraction \cite{frohl52,a:Krull}. Similarly, Fig.~\ref{fig:triplon_interaction_process}(b) illustrates that the creation of a pair of triplons, triplon propagation and subsequent pair annihilation induces a genuine three-triplon interaction. Besides pair creation and annihilation, i.e., the vacuum fluctuations, the hardcore property of triplons, excluding more than one per dimer, generates the interaction involving three triplons on three dimers. Otherwise the interaction would be a single-particle irreducible one, see Fig.~\ref{fig:triplon_interaction_process}(a). We stress that
this mechanism is not specific to one dimension, but applies in any dimension to
systems of coupled dimers \cite{cavad00a,quint12,McClarty2017} and, even more generally, to any system 
with finite Hilbert space dimension at each site because the latter implies 
a hardcore property, i.e., for spin flips.

To obtain a quantitatively correct description of the elementary excitations in the spin ladder we express the original model
\eqref{Jmodel} in terms of triplet creation and
annihilation operators, see Methods. Then, by a systematically
controlled change of basis, the model
can be mapped to one in which the number of triplons is conserved.
This is a vital step to make further analysis feasible and it
is achieved by means of a CUT
\cite{a:Wegner,a:Knetter,b:Kehrein,a:Fischer,a:Krull}.
The CUT is defined by the infinitesimal generator $\eta$ which
is classified by a label indicating whether the 2-triplon subspace is separated from  all subspaces with $n>2$ triplons (label ($2$:$n$))
or even the 3-triplon subspace (label ($3$:$n$)) is separated as well.
The resulting set of differential equations describes the
renormalization flow of the dispersion and the interactions.
These differential equations are truncated in some order in $x$
to keep their number finite and finally solved numerically.
In the course of the CUT the interactions of the triplons 
are renormalized and novel types of interactions are generated.
In particular, the two-triplon interaction is modified and three-triplon
interactions are induced. This is the mechanism how a genuine, irreducible three-body interaction comes into play. The possibility to distinguish 
the influence of the genuine two-triplon interactions from the one of 
the genuine three-triplon interactions is the crucial ace of the
CUT approach. An extensive derivation of the applied methods can be found in the Supplementary Methods.

\subsection*{Theoretical description of RIXS response}

To identify the effect of three-triplon interactions, the symmetries of 
spin ladders help greatly: reflection about the centerline defines
even and odd parity and the spin isotropic model conserves the total
spin. Since triplons are of odd parity \cite{trebs00,PhysRevLett.87.167204,schmi05b}
the excitations in the odd channel consist of one or three or five triplons and so on.
If in addition one uses a spin conserving (SC) 
probe the total spin of the excited states
is zero as in the ground state. A triplon has spin $S=1$, hence in the 
odd $S=0$ channel the leading contribution is the one formed by three triplons
only. This channel can be addressed by light scattering, for example
RIXS \cite{RevModPhys.83.705},  
or absorption in the terahertz (THz) range  \cite{Loidl2018N,PhysRevLett.123.067202}.
RIXS has the advantage of substantial momentum transfer due to the high energy of the scattered photons, in particular for hard X-rays as used for the Cu $K$-edge, while THz absorption provides high energy resolution, but stays at zero momentum. RIXS spectra can be resolved into spin-conserving ($\Delta S =0$) and spin non-conserving ($\Delta S \neq 0$) at the Cu $L_3$-edge of cuprates~\cite{PhysRevLett.112.147401, PhysRevB.99.134517}.
The spin-conserving channel can be well captured by the dynamical exchange structure factor (DESF). In contrast to the Cu $L$-edge, the channels at the Cu $K$-edge and oxygen $K$-edge in cuprates are purely spin conserving~\cite{PhysRevB.81.085124, Schlappa2018}, which is advantageous for experimental verifications of our predictions. RIXS resolution for hard X-rays has also improved considerably in the recent past~\cite{Ketenoglu:hf5287, Kim2018}.

To compute the RIXS response at the Cu $K$- and $L$-edge of cuprates, 
we use the dynamical correlation functions given by the 
ultra-fast core-hole liftime (UCL) approximation~\cite{PhysRevX.6.021020,PhysRevB.106.L060406}. 
We focus on the SC channel, not
accessible by inelastic neutron scattering which measures the dynamics structure factor
(DSF). For data of the DSF see Supplementary Note 1.
The SC channel is captured by the DESF given by
\begin{equation}
S^{\mathrm{ex}} (\mathbf{q}, \omega)  = \frac{1}{N} \sum\limits_{f} | \langle f | 
\sum\limits_{i,\tau} e^{\text{i} \mathbf{q} \mathbf{R}_{i,\tau}} O_{i,\tau}^\mathrm{ex} 
| g \rangle |^2  \delta ( E_f - E_g + \omega) 
\end{equation}
where $O_{i,\tau}^\mathrm{ex} = \mathbf{S}_{i,\tau} \cdot \left[ J_\text{leg} 
\left( \mathbf{S}_{i+1,\tau} + \mathbf{S}_{i-1,\tau}  \right) + J_\text{rung} 
\mathbf{S}_{i,\bar{\tau}} \right]$ (with $\bar\tau=1-\tau$) 
is the spin exchange observable,  
$|g\rangle$  and $|f\rangle$ are the ground and final states with energies 
$E_g$ and $E_f$, respectively, and $\omega$ is the energy loss to the system. 
Usually, the DESF is evaluated by exact diagonalization (ED)
\cite{PhysRevB.85.064423, PhysRevX.6.021020, PhysRevLett.103.047401, Kumar_2018} 
or by density matrix
renormalization group \cite{Nocera2018}.
These approaches provide spectra consisting of a multitude of finite-size peaks.
But they do not allow to trace back the physical origin of the spectral features.
The distinction between continuous scattering contributions and 
peaks from bound states can also be challenging.

Scattering states of two or more triplons lead to a continuous 
contribution to the DESF at fixed total momentum transfer
$\mathbf{q}$. This holds also for a two-triplon bound state scattering
with a single triplon. Only a bound three-triplon state yields an 
infinitely sharp $\delta$-peak in $S^{\mathrm{ex}} (\mathbf{q}, \omega)$
at given $\mathbf{q}$. Thus this is the smoking gun feature we have
to look for. For rendering purposes we will broaden it slightly;
but in experiments it will show up as sharp peak
limited only by the resolution of the apparatus.

\subsection*{Three-triplon bound states}

\begin{figure*}[htb]
	\includegraphics[width=1.00\textwidth]{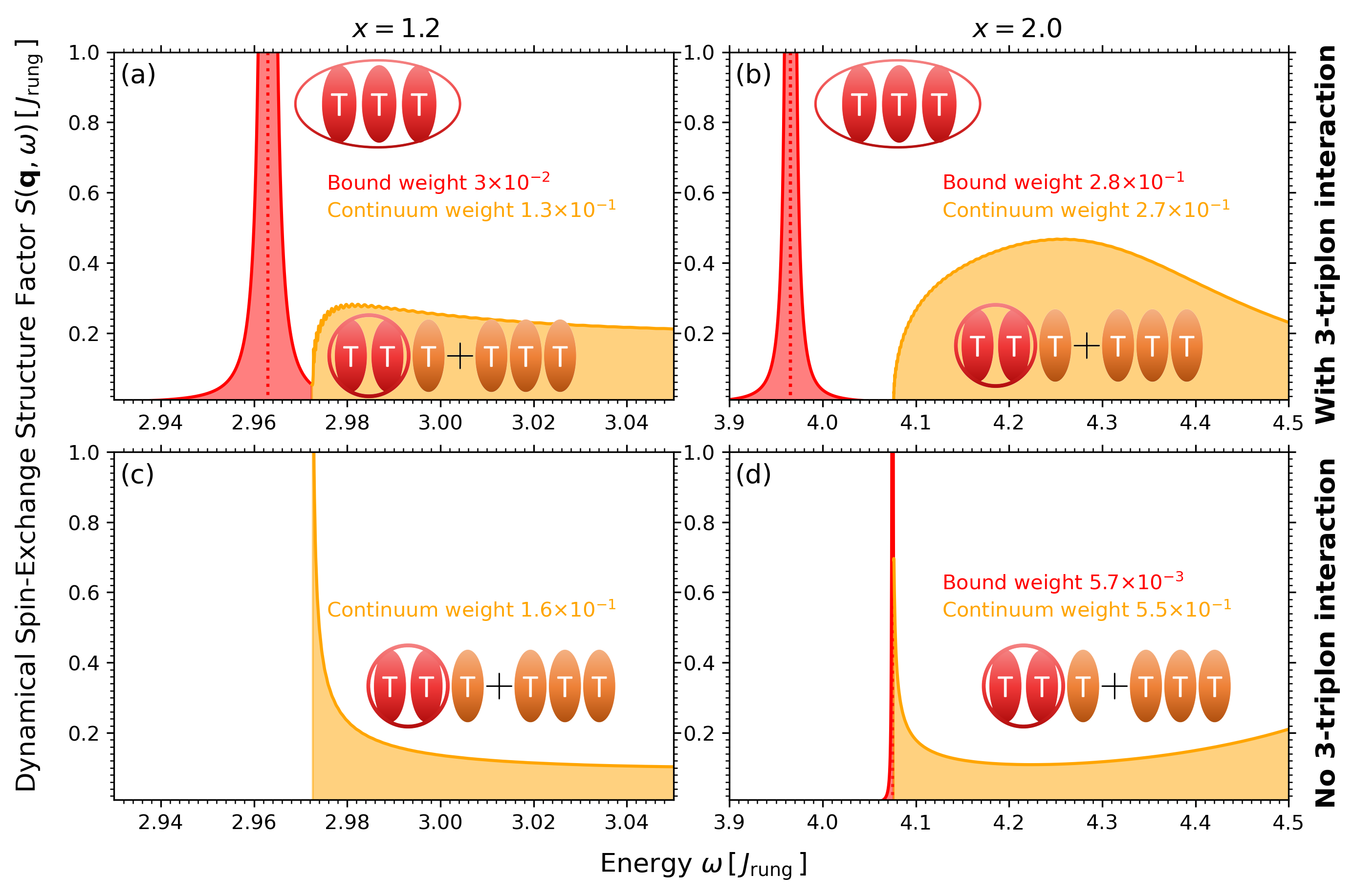}
	\caption{RIXS response in the $S=0$ channel for 
	$(q_x,q_y)=(\pi/(2a),\pi/a)$. The groups of three ellipses with the letter `T' denote three-triplon states. Asymptotically scattering triplons are colored orange and bound triplons are colored red. Panels (a) and (b) result from the mixed deepCUT calculations:	
	($2$:$n$)-generator in order $10$ for the single- and 
	two-triplon matrix elements, 	($3$:$n$)-generator in order $\ge 5$ for 
	the three-triplon matrix elements (see Supplementary Methods). 
	The effective observable was 	computed with the ($3$:$n$) generator. Panels 
	(c)-(d) results from the ($2$:$n$) 
	deepCUT calculations in order $10$ excluding irreducible three-triplon  interactions. 
	Panels (a) and (c) depict a system with $x=1.2$; 
	panels (b) and (d) with $x=2.0$.}
	\label{fig:Int3QP_Slices}
\end{figure*}

We provide data for $x=1.2$ and $x=2$ because these values
represent the experimentally relevant range in cuprates \cite{PhysRevLett.103.047401, UKumar2019, PhysRevLett.87.127002}
and the telephone number ladder La$_{5.2}$Ca$_{8.8}$Cu$_{24}$O$_{41}$ in particular.
Figure \ref{fig:Int3QP_Slices} depicts the DESF at $(q_x,q_y)=(\pi/(2a),\pi/a)$. 
Panels (a) and (b) show the three-triplon bound states that appear separated 
below the  three-triplon continuum for $x=1.2$ and $x=2.0$. The DESF of the 
bound state is plotted in red and the continuous DESF in orange. The latter stems
from three asymptotically free triplons and from scattering of a 
$S=1$ two-triplon bound  state with a single triplon.

We highlight that the three-triplon bound states have significant weight 
compared to the continua: for $x=2.0$ over $50\%$ of the spectral weight
resides in the three-triplon bound state. The energy separation between the bound state and 
continuum is small  for $x=1.2$ while it is becomes substantial for $x=2$. 
This makes  compounds with higher ratios $x$
interesting for the experimental verification of bound three-triplon states. For reference, CaCu$_2$O$_3$ is known to be a host for a spin ladder with large ratio $x$~\cite{Lake2010}.

To explicitly show the effect of the genuine three-triplon interactions
panels (c) and (d) of Fig.\ \ref{fig:Int3QP_Slices} display the DESF but without the three-triplon interactions. Then,
only a square-root divergence of the DESF appears at the lower band-edge.
For $x=2.0$ there may exist an extremely weakly bound state. But
the distinctive three-triplon bound states, clearly separated from the 
continuum only occurs as a consequence of the irreducible three-triplon interactions.
This is the key result of our investigation.

\begin{figure*}[htb]
	\includegraphics[width=1.00\textwidth]{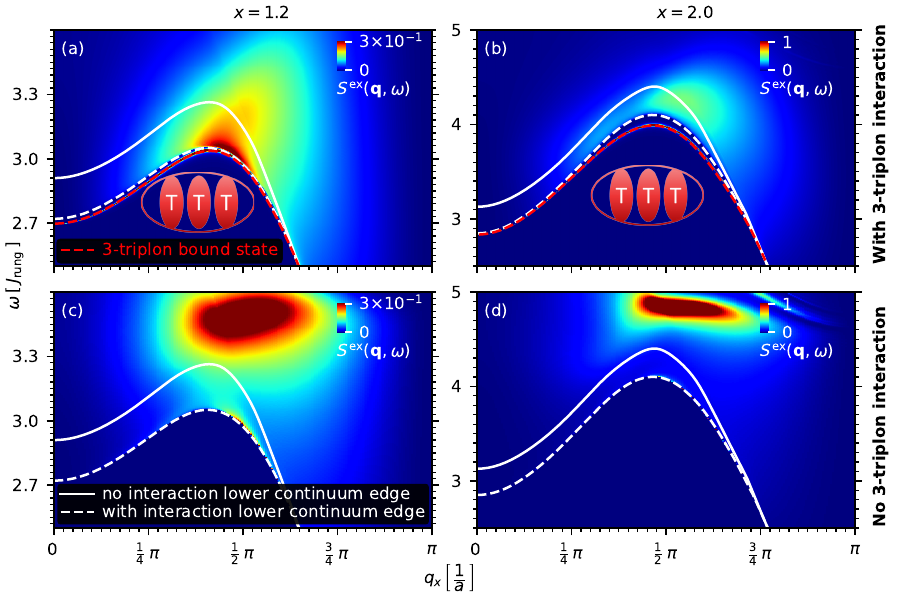}
	\caption{RIXS response in the $S=0$ channel for $q_y = \pi/a$. Three red ellipses with the letter `T' denote bound three-triplon bound states. Panels (a) and (b)
	result from the mixed deepCUT calculations, ($2$:$n$) generator in order $10$ for the
	single- and two-triplon matrix elements, ($3$:$n$)-generator in order $\ge 5$ for the 
	three-triplon matrix elements  (see Supplementary Methods). 
	The effective observable 	was computed with the ($3$:$n$)-generator. Panels (c) and 
	(d) results from the ($2$:$n$) deepCUT calculations in order $10$ excluding 
	irreducible three-triplon interactions. Panels (a) and 
	(c) depict data for $x=1.2$ while (b) and (d) for $x=2.0$.}
	\label{fig:Int3QP_Heatmap}
\end{figure*}

Figure \ref{fig:Int3QP_Heatmap} depicts the DESF at $q_y=\pi/a$ 
as function of $q_x$. Because of the symmetry 
$S(q_x,q_y,\omega)=S(-q_x,q_y,\omega)$ it is sufficient to show the
positive half of the Brillouin zone. In all cases, the total weight
is largest around $q_x=\pi/(2a)$.  While Fig.~\ref{fig:Int3QP_Heatmap} (a) and (b) depict the
full calculation with all interactions, Fig.~\ref{fig:Int3QP_Heatmap} (c) and (d) exclude 
the irreducible three-triplon interactions. 
The solid white line shows the lower edge of the three-triplon scattering 
continuum if only the one-triplon dispersion is considered, i.e.,
three asymptotically free triplons are involved.
The dashed white lines depict the actual lower continuum edge 
including scattering states from an $S=1$ two-triplon bound state and a 
single triplon. Hence the difference between the solid and the dashed line
for $q_x< 0.65\pi$ ($x=1.2$) and $q_x<0.75\pi$ ($x=2.0$), respectively,
stems from the irreducible two-triplon interactions. We emphasize
that only binding can induce states at lower energies. Without
binding, interactions can shift spectral weight, but only between the
edges of the continua. This is indeed one effect of the irreducible three-triplon 
interactions: they shift the weight significantly 
towards the lower band edge as can be clearly discerned by comparing Fig. \ref{fig:Int3QP_Heatmap} (a) with Fig. \ref{fig:Int3QP_Heatmap} (c) and Fig. \ref{fig:Int3QP_Heatmap} (b) with Fig. \ref{fig:Int3QP_Heatmap} (d).

The energy of the bound state formed from three triplons 
lies below the lower continuum edge for $q_x<0.65\pi/a$ ($x=1.2$) 
and $q_x<0.75\pi/a$ ($x=2.0$), respectively. These bound states are close 
to the lower continuum edges as expected from Fig.\ \ref{fig:Int3QP_Slices}. 
Their dependences on $q_x$  have a similar shape. 

\begin{figure}
	\centering
	\includegraphics[width=0.5\linewidth]{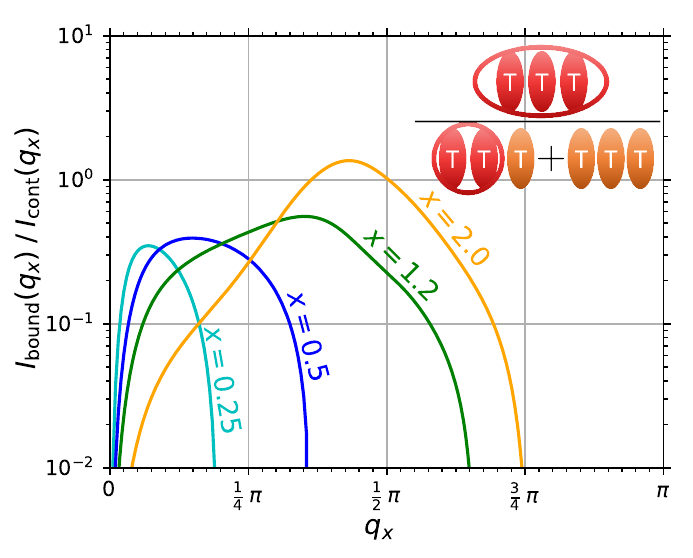}
	\caption{Ratio of bound to continuum excitations weight. The ratio is computed using the spectral weight  $I_\text{bound}$
	of the three-triplon bound state and the weight
	$I_\text{cont}$ of the continuum of three-triplon scattering states
	for $q_y = \pi/a$. The groups of ellipses with the letter `T' denote three-triplon states. Asymptotically scattering triplons are colored orange and bound triplons are colored red.}
	\label{fig:I_ratio}
\end{figure}

Figure \ref{fig:Int3QP_Slices} indicated that the spectral weight
of the bound state is significant relative to the weight in the adjacent continuum.
This feature holds generally for most values of $q_x$ as shown comprehensively 
in Fig.\ \ref{fig:I_ratio} for four different values of $x$ as function of $q_x$. 
The ratio of the spectral weight of the bound state to the spectral 
weight of all scattering states $I_{\text{bound}}/I_{\text{cont}}$ is plotted. 
We stress that the weight of the scattering states is integrated
up to the highest energies far beyond the energy of the three-triplon bound state. 
The maximum of the relative spectral weight moves to higher $q_x$ upon
increasing $x$, i.e., increasing $J_\text{leg}$. Also the relative weight 
increases with increasing $x$. Even for low values of $x$, the maximum of the 
weight of the bound state is never  an order of magnitude below
the weight in the continuum. For $x=2.0$, the bound weights
even exceed the continuum weights.

For completeness, the Supplementary Note 1 provides further 
data for the DESF and Supplementary Note 2 provides an estimate of the
importance of the contributions of four triplons and 
a comparison to data for finite
spin ladders obtained by ED. Comparison to the sum rule obtained from ED also shows, that the three-triplon sector contains almost all of the spectral weight for all $x$ considered here. Additionally, data for 
the standard dynamic structure factor is shown which is 
relevant for inelastic neutron scattering.
	
\section*{Discussion}

Binding effects are fundamental to understand the structure of
the surrounding matter ranging in length scale from femtometers to meters.
Most bound states stem from two-body interactions such as the Coulomb
potential. But in the present study, we have shown that in a realistic, correlated
condensed matter system genuine (irreducible) three-body interactions between the
elementary excitations are crucial as well. This is achieved for a generic
antiferromagnetic spin ladder as it is realized in cuprates well-known
from the field of high-temperature superconductivity, but the underlying
mechanism applies to wide classes of lattice systems in any dimension.

We systematically derived the three-body interactions of the elementary 
triplons by CUTs. We identified 
a probe channel (odd parity and $S=0$)
in which the three-triplon states represent the leading contribution.
We established that this channel can be probed by RIXS and
computed the DESF relevant for RIXS.
A particular asset of the CUT approach is that one can switch on or off the
genuine three-body interactions. In this way, we showed that it is the three-triplon
interactions which induce a significant shift of spectral weight in the DESF
to lower energies. Most notably, a bound
state formed from three triplons appears in a large part of the Brillouin zone 
-- but only if the three-triplon
interaction is taken into account. Although it is close in energy 
to three-triplon scattering states its weight is significant and partly 
dominates over the weight of the scattering states.

We highlight the fundamental difference of our finding to 
completely frustrating, antiferromagnetic diagonal couplings in spin ladders, where the number of triplets is already conserved so that triplets and triplons coincide. The complete frustration leads to an enhanced attraction of adjacent pairs of triplons such that three-triplon bound states (Fig.~\ref{fig:triplon_strings}(e)) 
and even states of $n$-triplon strings 
(Fig.~\ref{fig:triplon_strings}(f)) are predicted \cite{PhysRevB.94.094402}.
These local bound states stem from two-triplon interactions;
no three-body interactions are involved. Our key observation for unfrustrated spin ladders is that even for small $x$ the three-triplon interaction is sufficiently strong for the formation of a three-triplon bound state with small energy separation from the continuum. For $x \gtrsim 1$ this state gains significant spectral weight with a clear energy separation.

These results suggest that an experimental detection of the 
effects of three-body interactions is possible. Next-generation RIXS facilities 
with improved energy resolution should be able to tackle this challenge. Most promising candidate materials should have an intermediate to large leg over rung coupling $x$, such as  La$_{5.2}$Ca$_{8.8}$Cu$_{24}$O$_{41}$ \cite{PhysRevLett.87.127002},  CaCu$_2$O$_3$ \cite{Lake2010}, and La$_4$Sr$_{10}$Cu$_{24}$O$_{41}$ \cite{PhysRevLett.98.027403}, with the strongest response at $q_x=\pi/(2a)$ in the Brillouin zone. 
Alternatively, we point out that verification is also
conceivable by THz absorption which has a very high resolution in energy.
Here the challenge is that THz light hardly has any momentum $\hbar k$ so that only states with zero momentum are probed where no significant effects of the three-triplon interactions appear. But if a spin ladder material showed a slight distortion with periodicity of four rungs, the states at $q_x=m\pi/(4a)$ with $m\in\{1,2,3\}$ are folded to the center of the Brillouin zone and three-triplon binding would become detectable \cite{PhysRevB.93.241109,C4DT01746C}, as for instance shown for screwed spin chains \cite{Loidl2018N}. In these systems, the oxygen $K$-edge, which probes spin-conserving channel exclusively, can also be used in spite of it having access to restricted momentum transfer~\cite{Schlappa2018, Kumar_2018}.

The theoretical results open up the stage for the study of the effects of genuine three-body interactions occurring in generic lattice models for condensed matter. 
In one dimension in particular, where quantum effects are typically strongest, 
they could be an important ingredient for the formation of $n$-triplon strings (Fig.~\ref{fig:triplon_strings}(f)) generalizing the previously detected Bethe strings \cite{Loidl2018N,PhysRevLett.123.067202} to a much broader class of solid state systems not restricted to integrable systems such as spin chains. Our results open perspectives for ways to investigate complex bound states, connecting to various cross-boundary efforts with the same aim, such as the discovery of three-body correlations in ultracold atoms \cite{2017Sci355377F}, the formation of topological bound states in non-Hermitian systems of photonic lattices \cite{Weimann2017} and bound states for quantum computation in superconducting nanowires \cite{Higginbotham2015}. With recent improvements in experimental techniques, we are hopeful of verification of our findings in the near future.

\section*{Methods}
\textsl{Further details on the methods and numerical details can be found in the Supplementary Methods.}

\subsection*{Triplet and Triplon representation}
Starting from the regime of strong rung couplings, we reformulate \eqref{Jmodel} in terms of triplet
creation operators $t_i^{\alpha,\dag} |s\rangle = |t^\alpha_i\rangle$ where $|s\rangle$ is the singlet 
ground state for $x=0$ and $|t^\alpha_i\rangle$ is an $S=1$ triplet state of flavor $\alpha \in \lbrace x,y,z \rbrace$ 
on rung $i$. The triplet annihilation operator $t_i^{\alpha}$ annihilates the specific triplet at rung $i$ and
restores the singlet on this rung.
The triplet operators satisfy the hardcore boson algebra 
$\left[ t^\alpha_i, t^{\beta\,\dagger}_j \right] = \delta_{i,j} \left( \delta_{\alpha,\beta} 
\left( \mathds{1} - \sum_{\gamma\in\{x,y,z\}} t^{\gamma \,\dagger}_i t^\gamma_i \right) 
- t^{\beta \,\dagger}_i t^\alpha_i \right)$. 
Representing the spin operators by the triplet operators allows us to map the spin systems onto an 
interacting quasiparticle problem \cite{a:Krull}. In the course of the CUT, the triplet operators are 
gradually converted to conserved triplon operators.
Note that we distinguish between unconserved triplets and conserved triplons 
here. This distinction is not always utilized in literature.
Explicit expressions for the Hamiltonian and the observables in the triplet language are given in the Supplementary Methods. 

\subsection*{Foundation of CUT}

The basic idea of a dedicated unitary transformation is to change the basis such that the problem under study
becomes easier to tackle. The continuous transformation has the advantage to perform the basis change
in a renormalizing fashion, i.e., processes linking large energy difference are transformed most quickly
and processes with smaller and smaller energy differences are eliminated slower and slower in the flow of the renormalization parameter $\ell\to\infty$. This is important because one can never track the complete
Hamiltonian along the unitary flow, but truncations are necessary. The renormalizing property
helps to obtain robust effective models \cite{a:Wegner,kehre96b,b:Kehrein,a:Krull,PhysRevB.87.184406,PhysRevB.94.180404}.

We use CUTs here to eliminate the terms in the Hamiltonian which 
change the number of particles, i.e., the number of triplons. 
In this way, this number becomes a conserved quantity, the ground state is the vacuum of triplons and
hence the calculations of dynamic zero-temperature correlations are simplified to tractable
few-particle calculations. To evaluate the spin conserving contributions to the RIXS spectra 
we decouple the interacting multi-triplon sectors from each other by applying the CUT to the 
Hamiltonian and to the relevant observables. The resulting effective Hamiltonian consists of 
$n$-particle irreducible parts, i.e., the one-triplon dispersion, the two-triplon interaction, 
and the three-triplon interactions which are of particular interest here. Irreducible higher 
triplon interactions, e.g., of four triplons, also occur, but 
only matter if four or more triplons are considered.

\subsection*{CUT generator scheme}

The CUT provides a set of differential equations for the prefactors of a large number
of interactions which proliferate exponentially with system size. Hence truncations
are required which systematically control the accuracy of the obtained 
effective Hamiltonian and effective observables. Here we employ two concepts.
First, we do not separate all $m$-triplon states from all $n$-triplon states
with $m\ne n$, but use either the ($2$:$n$) or the ($3$:$n$) generator. The generator
($m$:$n$) separates the $0, 1, \ldots m$ triplon states from all higher states
with $n>m$ \cite{a:Fischer}, i.e., we decouple either the zero, one, and two triplon states
from the rest or we decouple the zero, one, two, and three triplon
states from the remaining states. The latter, however, does not work up to very high 
order in $x$, see below.

Second, we use the expansion parameter $x=J_\text{leg}/J_\text{rung}$
to classify the various processes. We keep all processes which influence
the effective Hamiltonian up to a certain order in $x$. The physics behind this idea
is that a larger order translates to a larger range in space, i.e., to a larger
distance between the interacting triplons on the rungs. This concept leads
to large, but tractable sets of differential equations which then are integrated
numerically. The whole procedure is dubbed directly evaluated enhanced perturbative
CUT or deepCUT for short and has proven to be reliable and efficient\cite{a:Krull} up 
to $x=3$.  The ($3$:$n$)-generator was used to calculate the 
irreducible three-triplon contributions. If the highest robust order was smaller 
than 10, 
the ($2$:$n$)-generator was used to calculate the irreducible two-triplon and 
one-triplon contributions 
in order 10. 

The accuracy of the CUT approach is examined by comparing it to lower order CUT calculations 
as well as to results from ED on finite systems in the Supplementary Note 2. 

\subsection*{Details on dynamic structure factors}

Based on the effective observables obtained from the CUT we use Lanczos tridiagonalization to compute the 
continued fraction expansion of the DSF and 
the DESF, see Supplementary Methods. This renders finite-size effects negligible.
At the same time we can identify the physical processes that contribute to the spectral functions. 
A square-root terminator was employed to terminate the continued fraction so that the contribution
of the scattering states is found as continuous function without resorting to artificial broadening \cite{petti85}.

The bound states are mathematically $\delta$-distributions in the spectral densities, i.e., they
are infinitely sharp, which prevents a graphical representation. Thus,  the bound states (and only them) 
are artifically
broadened in our plots. They are displayed as Lorentzians of width $5\cdot 10^{-4}\,J_{\text{rung}}$. 

The lower edge of the three-triplon scattering states including two-triplon interactions is calculated 
from the converging  values of the Lanczos coefficients \cite{petti85}. 
The lower edge at wave number $q$ without any interactions is calculated directly from 
$\text{min}_{q=k_1+k_2+k_3} (\omega(k_1)+\omega(k_2)+\omega(k_3))$ where $\omega(k)$ is the
triplon dispersion.

\section*{Acknowledgements}

This work was supported by the U.S. DOE NNSA under Contract No. 89233218CNA000001 
via the LANL LDRD Program. G.S.U. is supported by the DFG under project No. UH90/14-1.

\newpage

\end{document}